\documentclass[prc,lengthcheck,showpacs,superscriptaddress]{revtex4}

\usepackage{graphicx,color}
\usepackage{dcolumn}

\begin{document}

\preprint{TRI-PP-02-18}

\title{Parity Violation in Proton-Proton Scattering at 221 MeV}

\author{A.~R.~Berdoz}
\author{J.~Birchall}
\author{J.~B.~Bland}
\affiliation{Department of Physics and Astronomy, University of Manitoba,
Winnipeg, MB, Canada R3T 2N2}
\author{J.~D.~Bowman}
\affiliation{Physics Division, Los Alamos National Laboratory,
Los Alamos, NM 87545, USA}
\author{J.~R.~Campbell}
\affiliation{Department of Physics and Astronomy, University of Manitoba,
Winnipeg, MB, Canada R3T 2N2}
\author{G.~H.~Coombes}
\affiliation{TRIUMF, 4004 Wesbrook Mall, Vancouver, BC, Canada V6T 2A3}
\author{C.~A.~Davis}
\affiliation{Department of Physics and Astronomy, University of Manitoba,
Winnipeg, MB, Canada R3T 2N2}
\affiliation{TRIUMF, 4004 Wesbrook Mall, Vancouver, BC, Canada V6T 2A3}
\author{A.~A.~Green}
\affiliation{Department of Physics and Astronomy, University of Manitoba,
Winnipeg, MB, Canada R3T 2N2}
\author{P.~W.~Green}
\affiliation{Centre for Subatomic Research, University of Alberta,
Edmonton, AB, Canada T6G 2N5}
\author{A.~A.~Hamian}
\affiliation{Department of Physics and Astronomy, University of Manitoba,
Winnipeg, MB, Canada R3T 2N2}
\author{R.~Helmer}
\author{S.~Kadantsev}
\affiliation{TRIUMF, 4004 Wesbrook Mall, Vancouver, BC, Canada V6T 2A3}
\author{Y.~Kuznetsov}
\thanks{deceased}
\affiliation{Institute for Nuclear Research, Academy of Sciences of Russia,
RU-117334 Moscow, Russia}
\author{L.~Lee}
\affiliation{Department of Physics and Astronomy, University of Manitoba,
Winnipeg, MB, Canada R3T 2N2}
\author{C.~D.~P.~Levy}
\affiliation{TRIUMF, 4004 Wesbrook Mall, Vancouver, BC, Canada V6T 2A3}
\author{R.~E.~Mischke}
\affiliation{Physics Division, Los Alamos National Laboratory,
Los Alamos, NM 87545, USA}
\author{N.T.~Okumusoglu}
\affiliation{Rize Faculty of Sciences and Arts, Karadeniz Technical
University, Rize, Turkey}
\author{S.~A.~Page}
\author{W.~D.~Ramsay}
\author{S.~D.~Reitzner}
\affiliation{Department of Physics and Astronomy, University of Manitoba,
Winnipeg, MB, Canada R3T 2N2}
\author{T.~Ries}
\affiliation{TRIUMF, 4004 Wesbrook Mall, Vancouver, BC, Canada V6T 2A3}
\author{G.~Roy}
\affiliation{Centre for Subatomic Research, University of Alberta,
Edmonton, AB, Canada T6G 2N5}
\author{A.~M.~Sekulovich}
\affiliation{Department of Physics and Astronomy, University of Manitoba,
Winnipeg, MB, Canada R3T 2N2}
\author{J.~Soukup}
\author{G.~M.~Stinson}
\author{T.J.~Stocki}
\affiliation{Centre for Subatomic Research, University of Alberta,
Edmonton, AB, Canada T6G 2N5}
\author{V.~Sum}
\affiliation{Department of Physics and Astronomy, University of Manitoba,
Winnipeg, MB, Canada R3T 2N2}
\author{N.~A.~Titov}
\affiliation{Institute for Nuclear Research, Academy of Sciences of Russia,
RU-117334 Moscow, Russia}
\author{W.~T.~H.~van Oers}
\author{R.~J.~Woo}
\affiliation{Department of Physics and Astronomy, University of Manitoba,
Winnipeg, MB, Canada R3T 2N2}
\author{S.~Zadorozny}
\author{A.~N.~Zelenski}
\affiliation{Institute for Nuclear Research, Academy of Sciences of Russia,
RU-117334 Moscow, Russia}

\collaboration{The TRIUMF E497 Collaboration}
\noaffiliation

\date{\today}

\begin{abstract}

TRIUMF experiment 497 has measured the parity violating longitudinal analyzing
power, $A_z$, in $\vec{p}p$ elastic scattering at 221.3 MeV incident proton
energy. This comprehensive paper includes details of the corrections, some of
magnitude comparable to $A_z$ itself, required to arrive at the final result. 
The largest correction was for the effects of first moments of transverse
polarization. The addition of the result, $A_z=(0.84 \pm 0.29 (stat.) \pm 0.17
(syst.)) \times 10^{-7}$, to the $\vec{p}p$ parity violation experimental data
base greatly improves the experimental constraints on the weak meson-nucleon
coupling constants $h^{pp}_\rho$ and $h^{pp}_\omega$, and also has implications for
the interpretation of electron parity violation experiments.

\end{abstract}


\maketitle

\section{Introduction}

This experiment determines the parity-violating longitudinal analyzing
power, $A_z = (\sigma^+ - \sigma^-)/(\sigma^+ + \sigma^-)$, in $\vec{p}p$
elastic scattering, where $\sigma^+$ and $\sigma^-$ are the scattering
cross sections for positive and negative helicity. The measurements were
performed in transmission geometry, with beam energy and detector
geometries selected to ensure that parity mixing in the lowest order
$^1S_0-^3P_0$ partial wave amplitude did not contribute to the measured
$A_z$, hence leaving a parity violating asymmetry arising almost entirely
from $^3P_2-^1D_2$ mixing \cite{Sim88}. This amplitude has not previously
been studied experimentally, and the possibility is unique to the energy
regime accessible with the TRIUMF cyclotron. The energy at which the
contribution to $A_z$ from the lowest order $^1S_0-^3P_0$ mixing vanishes
is determined by the well known strong nuclear phase shifts. The scale
factors multiplying this and other partial wave contributions are set by
the weak interaction. $\vec{p}p$ parity violation experiments determine
these scale factors experimentally. In the context of the weak meson
exchange model \cite{DDH80}, the TRIUMF measurement of $A_z$ determines
primarily the weak $\rho$-meson-nucleon coupling constant $h^{pp}_\rho$ =
($h^{0}_{\rho}+h^{1}_{\rho}+h^{2}_{\rho}/\sqrt 6$), where the superscripts
refer to isospin change \cite{Sim7581}. Precision results already obtained
by the SIN group at 45 MeV \cite{Kist87} and the Bonn group at 13.6 MeV
\cite{Evers} determined essentially the sum $h^{pp}_\rho + h^{pp}_\omega$,
where $h^{pp}_\omega = h^{0}_{\omega}+h^{1}_{\omega}$. With the addition of
the TRIUMF result at 221.3 MeV, $h^{pp}_\rho$ and $h^{pp}_\omega$ are now
determined separately for the first time.

\begin{figure}[tb]
\includegraphics[width=\linewidth]{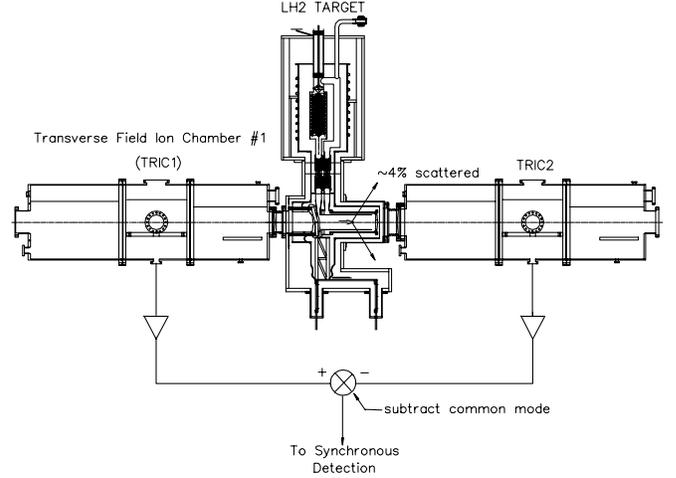}

\caption{Principle of the measurement. A longitudinally polarized proton beam
was passed through a liquid hydrogen target. The beam current before and after
target was detected and the signals were subtracted. The presence of a
component of the difference signal synchronized with spin flip indicates a
parity violating dependence of the transmission through the target on the
helicity of the incident protons.}

\label{principle}
\end{figure}

It is important to have experimentally determined values of the weak
meson-nucleon couplings, as theoretical calculations of their values are
quite uncertain and the correct values are helpful in the interpretation of
the results of other parity violation experiments. They are needed, for
example, in calculations of the proton anapole moment \cite{Zhu00}, one of
the radiative corrections to the electron-nucleon isovector axial form
factor in experiments such as SAMPLE \cite{Hast00} and G$^0$ \cite{gzero}.

\section{Background}

TRIUMF E497 \cite{Page87}, was first funded in 1988, and the required new
beamline was completed in 1994. A major effort to understand and minimize
systematic error contributions was then undertaken. Following many years of
effort that resulted in the reduction of both the helicity correlated beam
modulations $\Delta x_i$ and the sensitivities $\frac{\partial
A_z}{\partial x_i}$, to these modulations, the first significant data set
for E497 was acquired in February and March of 1997, with a statistical
error on $A_z$ of $\pm 0.5 \times 10^{-7}$, and most systematic
errors at or below the 10$^{-7}$ level. That result represented a major
milestone for the experiment \cite{Ham98}. Data taking continued in 1998
and 1999, the final re-analysis of all the data was completed in early 2001
\cite{Bland01}, and the result for $A_z$ was published \cite{Berd01b}. The
present paper presents detailed descriptions of the experiment, the data
analysis, and systematic error corrections, that were not included in the
Letter \cite{Berd01b}.

\begin{figure}[tb]
\includegraphics[width=\linewidth]{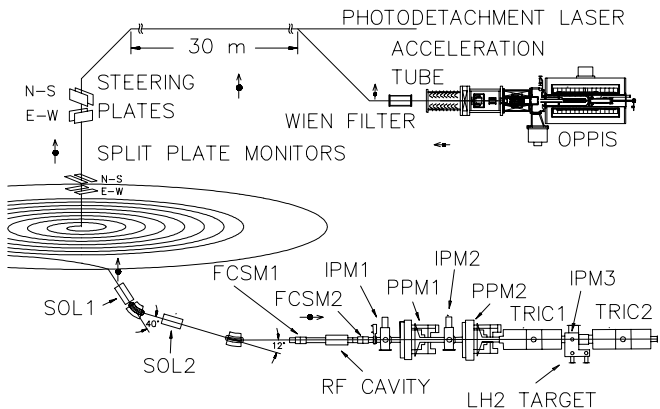}
\caption{General layout of the TRIUMF parity experiment.  (OPPIS:
Optically Pumped Polarized Ion Source;  SOL: Spin Precession Solenoid;
FCSM: Ferrite Cored Steering Magnet; IPM: Intensity Profile Monitor; PPM:
Polarization Profile Monitor; TRIC: Transverse Field Ionization Chamber)}
\label{general}
\end{figure}

\section{Experimental Setup}

\subsection{General}

\begin{figure}
\includegraphics[height=\linewidth, angle=90]{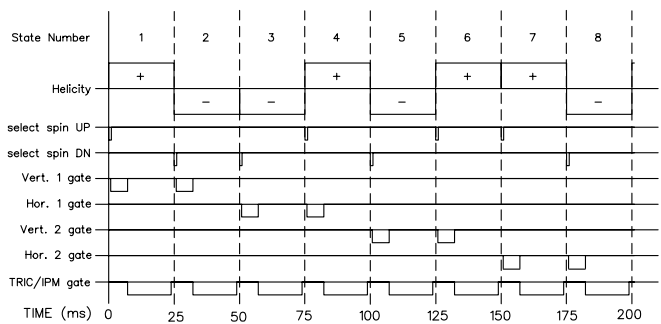}

\caption{Sequence of events in one 200 ms, 8-state cycle. Each state starts with
a PPM scan. In the first two states of an octet, PPM1 scans vertically, on the
next two states, PPM1 scans horizontally.  This 4-state sequence is then
repeated with PPM2. }

\label{timing}
\end{figure}

The principle of the experiment is straightforward. A longitudinally polarized
proton beam is passed through a liquid hydrogen target and the change in
transmission when the spin of the incident protons is reversed is the parity
violating signal. To measure this, the beam current before and after the liquid
hydrogen target was measured and the signals were subtracted (Fig.
\ref{principle}). The spin state was then changed in a special pattern at 40
spin states per second and synchronous detection was used to extract that
component of the difference signal that was synchronized with the spin flip.
Unfortunately, beam parameters other than spin changed when the spin was
flipped, and great pains had to be taken to measure, and correct for, the false
signals resulting from these unwanted helicity correlated changes. Technical
details of the systems required to do this will be described in detail in what
follows, but an idea of the complexity can be obtained from Fig. \ref{general},
identifying the major pieces of equipment, and Fig. \ref{timing}, showing the
sequencing of the control measurements within a 200 ms, eight spin-state
cycle. 

Because of the importance of understanding systematic errors in such an
experiment, essentially all the pieces of equipment, from the ion source to the
beam dump, had to be considered integral parts of the parity experiment. Figure
\ref{general} shows the main subsystems of the experiment -- the TRIUMF
optically pumped polarized ion source (OPPIS), the cyclotron, the beam
transport, and the specialized parity instrumentation \cite{vano99}. A 5 $\mu$A
transversely polarized beam was transported to the cyclotron through an
approximately 50 m long injection beamline.  A 200 nA beam at 75 - 80\%
vertical polarization was extracted at 221.3 MeV. Spin precession through pairs
of solenoid and dipole magnets resulted in delivery of a longitudinally
polarized beam to the parity apparatus. There were two complementary states of
the spin transport, the ``positive helicity'' and ``negative helicity''
beamline tunes, which transported spin-up in the cyclotron into either $+$ or
$-$ helicity at the parity apparatus.

In the last section of the beam line, the longitudinally polarized beam
first passed through a series of diagnostic devices -- two beam intensity
profile monitors (IPMs) and two transverse polarization profile monitors
(PPMs) -- before reaching the $LH_2$ target, which was preceded and
followed by transverse electric field, parallel plate ionization chambers
(TRICs) to measure the target transmission. A third IPM was located
immediately in front of the $LH_2$ target, inside the cryostat vacuum.

\subsection{Optically Pumped Ion Source}
\label{sec:oppis}
In the ion source, 9 W of 795 nm laser light was used to optically pump a
rubidium vapor whose polarization was ultimately transferred to the protons of
the $H^-$ beam. The polarization was reversed using small tilting etalons to
make rapid frequency adjustments to the two pumping lasers to match either the
$\sigma^+$ or $\sigma^-$ component of the rubidium D$_1$ transition; the two
are separated by 93.5 GHz. No macroscopic electric or magnetic fields were
altered. This minimized helicity correlated changes in accelerated beam
parameters other than polarization. This was very important to the success of
the experiment. The ion source is the ultimate origin of all helicity
correlated modulations, and the more that unwanted modulations could be reduced
at the source the less corrections were required later. The extracted beam
current required by the experiment was not large (200 nA) so, to reduce
unwanted helicity correlated modulations, it was possible to sacrifice most of
the OPPIS intensity in return for beam quality. For example, the RF bunchers in
the injection beamline, which enhance the cyclotron transmission by a factor of
five, also amplify the coherent energy modulations by two orders of magnitude,
and could not be used during the parity experiment. High-current OPPIS
development \cite{Zelen96, Zelen97, oppis} for high energy accelerators
proceeded at the same time as development of the TRIUMF source and contributed
greatly to the parity violation experiment.

The polarization of the rubidium vapor was monitored and controlled
on-line by observing the Faraday rotation of light from an additional
100 mW, TiS probe laser that emitted 780.8 nm light -- close to the
D$_2$ transition of rubidium. The polarization of the linearly polarized
probe laser light rotated through an angle proportional to the rubidium
vapor polarization. The Faraday measurements also provided confirmation of
the helicity state of OPPIS. The Faraday rotation signal was encoded as a
frequency to prevent helicity correlated signals from being present in the
electronics racks. Details of the OPPIS Faraday rotation system are
described elsewhere \cite{Reitz96}.

Every effort was made to tune OPPIS for minimum intensity change on
spin flip. The main technique used to do this was to keep the rubidium
polarization, as measured by the Faraday rotation, at close to 100\% and
with the two spin states matched to better than $\pm$0.5\%. It was not
possible to eliminate helicity correlated current modulation completely,
but under normal data taking conditions, helicity-correlated current
modulations $\Delta I / I = (I^+ - I^-)/(I^+ + I^-)$ of a few parts in
$10^5$ were routinely achieved.

To enable measurement of the sensitivity of the experiment to coherent
intensity modulation, provision was made at the ion source to produce an
intentional, controlled, intensity modulation when desired. This was done using
an auxiliary 18 W green argon laser (Spectra Physics Inc., operating on all
lines with most power at 514 nm and 488 nm) beam that co-propagated with the
$H^-$ beam along the 30 m long horizontal section of the injection beam line,
neutralizing through photodetachment a small fraction of the beam. The
photodetachment laser could be interrupted synchronously with the parity spin
sequence, so that the beam current in every other ``spin off'' (i.e. with the
optical pumping lasers blocked with a shutter) data taking cycle was modulated
at the 0.1\% level.

\subsection{Cyclotron and Beamline}

\subsubsection{Beam Transport}

After extraction from OPPIS, the $H^-$ beam passed through a Wien filter that
was tuned to produce vertical polarization at the entrance to the cyclotron.
The injection line from the ion source to the cyclotron used electrostatic
elements and was magnetically shielded from the fringe field of the cyclotron.
Instabilities in the polarized ion source, beamline power supplies, and
mechanical vibrations of the whole injector building structure, cause the beam
position at the injection point to fluctuate. These fluctuations are converted
to energy and current modulations of the accelerated beam. These were
significantly reduced by a position stabilization feedback system installed in
the vertical section of the injection beamline \cite{Zelen98}. The system was
based on two split-plate beam position monitors, with correction voltages
applied to electrostatic steering plates. About 50\% of the beam was lost on
the split-plates, but the result was a significant improvement in the beam
stability, and a reduction in noise. The sampling rate of the integrated
current feedback amplifier was 1 kHz, so spin-flip correlated position
modulations produced in the source at 40 Hz were also reduced by the position
stabilization system. 

After injection into the cyclotron, the beam was accelerated to 221.3 MeV in
the cyclotron, and was extracted by a thin stripping foil. Various stripping
foil designs were tried, but most of the data were taken with a 2.5 mm wide x
26 mm high x 5 mg/cm$^2$ thick, pyrolytic graphite foil. This foil was mounted
in a special ``bow-saw'' shaped holder that supported the foil from both ends
to prevent curling. Following extraction from the cyclotron, the proton
polarization vector was precessed through $\pm \sim 63^\circ$ in the first
solenoid, 88.61$^\circ$ in the first (40$^\circ$) dipole, $\pm \sim 87^\circ$
in the second solenoid, and 26.58$^\circ$ in the final (12$^\circ$) dipole,
resulting in a longitudinally polarized beam that was transported to the parity
apparatus. The sign of the solenoid rotation was chosen depending on the
desired helicity of the tune and the exact solenoid strengths were fine tuned
empirically to produce pure longitudinal polarization in the presence of the
cyclotron fringe field. In contrast with systems using only one solenoid and
one dipole, which can produce longitudinal polarization only at one energy, the
TRIUMF system is capable of producing longitudinal polarization at any proton
energy up to 500 MeV. \cite{longpol}. 

Ideally, the beam transport should produce an achromatic waist downstream of
the $LH_2$ target, to minimize the effect of first moments of transverse
polarization, which are the dominant source of systematic error in this
experiment, as they couple to the large ($\sim 0.3$) parity allowed transverse
analyzing power $A_y$. The rotation of phase space introduced by the
superconducting solenoid magnets was approximately compensated for by rotating
the quadrupole doublet before the final 12 degree bend by $37.4^\circ$ and the
quadrupole triplet following the bend by $17.2^\circ$. To reverse the spin
direction at the parity target relative to the spin in the cyclotron, the
directions of the solenoidal fields and the signs of the quadrupole rotations
were reversed. In principle, that should have been be all that was required. 
In practice, small adjustments to some quadrupole and steering magnets were
necessary, but empirical tunes were developed that kept these changes very
small (1-2\%).

Between the second solenoid magnet and the final $12^\circ$ bend, where the
polarization of the beam has both a longitudinal and a sideways component,
a four branch polarimeter \cite{ibp} was used to measure the transverse
polarization components at regular intervals. Since the angle of the
polarization vector is known at this location, the absolute beam
polarization could be determined. Once the beam passed the final dipole
magnet, the polarization was longitudinal and the PPMs determined the small
transverse components. The upstream polarimeter was also used, with purely
vertically polarized beam, to check the absolute calibration of the PPMs.

Just after the upstream polarimeter, a Beam Energy Monitor (BEM)
\cite{Gree81} was available to measure the energy of the proton beam. The
BEM achieves a statistical precision of $\pm$ 20 keV in approximately one
hour. The absolute calibration is known to $\pm 230$ keV \cite{Dav96}. The
(BEM) target could not be inserted during normal data taking, so periodic
BEM measures were made to check the beam energy. Table \ref{energies}
summarizes the beam energy measurements.

\begin{table}
\caption{Beam energy measurements made during the three running
periods.  The uncertainties shown are statistical only.  The absolute
calibration of the BEM is known to $\pm 0.23$ MeV.
}
\begin{tabular}{lr}
\toprule
 Running period  & beam energy   \\
\colrule
 1997   &  $221.30 \pm 0.03$ MeV  \\
 1998   &  $221.5 \pm 0.1$ MeV   \\
 1999   &  $221.3 \pm 0.1$ MeV   \\
\botrule
\end{tabular}
\label{energies}
\end{table}

\subsubsection{Beam Energy}
\label{subsec:energy}

The beam energy of 221.3 MeV at the target entrance was chosen so that the
$(^1S_0-^3P_0)$ contribution would be zero when the acceptance of the
detectors and the energy loss in the target was taken into account. This
was done using a Monte Carlo program that simulated the experiment using
detailed target and detector geometries, and including the energy
dependence of the strong $pp$ phase shifts. It was found that the beam
energy was not overly critical; a departure of one MeV from the energy for
which the $(^1S_0 - ^3P_0)$ contribution integrated to zero caused this
mixing to contribute only $\sim 10^{-9}$ to $A_z$. Theoretical calculations
of $A_z$ based on total cross section, and which assume no energy loss in
the target, show the $(^1S_0 - ^3P_0)$ zero at 225 MeV, where the $^1S_0$
and $^3P_0$ strong phase shifts are equal and opposite \cite{said}.
According to the Monte Carlo simulation, this theoretical $A_z$ at 225 MeV
should be compared to the TRIUMF $A_z$ multiplied by $1.02 \pm 0.02$.

\subsection{Specialized Instrumentation}
\label{sec:instrumentation}

\subsubsection{General}

Very strict constraints had to be imposed on the incident longitudinally
polarized beam to limit both random and helicity correlated modulations of
the beam intensity, energy, horizontal ($x$) and vertical ($y$) position and
direction, beam width ($\sigma_x$ and $\sigma_y$), transverse polarization
($P_x$ and $P_y$), and first moments of transverse polarization ($\langle
xP_y \rangle$ and $\langle yP_x \rangle$). The approach followed was to
design the experimental apparatus for minimum sensitivity to beam property
variations, to monitor helicity-correlated beam properties during data
taking, and to correct the data for all significant helicity correlated
effects. A program of auxiliary calibration measurements was interwoven in
the regular data taking cycle in order to establish the sensitivity of the
apparatus to all measurable systematic effects under data taking
conditions.

The custom-built parity instrumentation occupied approximately the last 10
m of the beamline, between the last quadrupole magnet and the west wall in
the TRIUMF proton hall (beamline 4A/2). Transverse field parallel plate ion
chambers TRIC1 and TRIC2  measured the beam current incident on and transmitted
through the target. The parity violation signal was derived from the
helicity-correlated analog signal difference between the beam currents
measured by the two TRICs. Upstream of the target were two polarization
profile monitors (PPMs) to measure the distributions of transverse
polarization $P_y(x)$ and $P_x(y)$ across the beam. The three intensity
profile monitors (IPMs), measured the intensity distribution of beam
current in $x$ and $y$. Two of the IPMs were coupled to a pair of fast
ferrite cored steering magnets (FCSMs) that locked the beam path on the
desired axis through the equipment.

\subsubsection{Intensity Profile Monitors}
\label{subsubsec:ipm}

The IPM signals were based on secondary emission from thin nickel foil strips
 arranged in harps  placed between 8 $\mu$m thick aluminum foils (A very similar,
earlier version is described in \cite{Berd91} and \cite{Sekul90}). For IPM1 and
IPM2 the nickel strips were 3 $\mu$m thick, 1.5 mm wide, and spaced 2.0 mm
center to center. In IPM3, the nickel strips were 10 $\mu$m thick, 2.5 mm wide,
and spaced 3.0 mm center to center. Each IPM contained a vertical
and a horizontal harp with 31 strips per harp.  IPM1 also contained
an aluminum normalization foil that provided a beam current signal to the
liquid hydrogen target controller. The 31 signals from the strips of each of
the six planes were individually amplified and digitized to provide the beam
intensity profiles in each spin state. Hardware beam centroid evaluators
delivered signals proportional to the beam centroids. Corresponding correction
signals were used to drive feedback loops to the pair of horizontal and
vertical fast FCSMs (Fig \ref{general}). This allowed the beam intensity
profile centroids to be kept within $\pm 1 \mu$m with a dc offset less than 50
$\mu$m from the desired axis. Sensitivities to helicity-correlated position and
size modulations were determined with the beam unpolarized and with enhanced
modulations introduced using the fast, FCSMs synchronized to the spin sequence.

\subsubsection{Polarization Profile Monitors}

\begin{figure} 
\includegraphics[width=40mm]{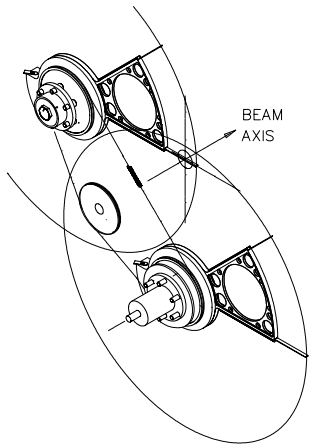}

\caption{Method of scanning the CH$_2$ blades through the beam. The blades
measure 5 mm x 1.6 mm in cross section and extend 85 mm beyond the holders.
Each PPM has two horizontally scanning and two vertically scanning blades. This
figure shows a blade at the middle of a vertical scan.}

\label{blades} 
\end{figure}

The PPMs are described in detail in \cite{Berd01}. Each PPM has four high
density polyethylene (CH$_2$) blades, 5 mm x 1.6 mm in cross section. The
blades were mounted on wheels rotating at 5 revolutions per second, and were
scanned through the beam as shown in Fig. \ref{blades}, two blades scanning
vertically through the beam and two blades scanning horizontally. The figure
shows one blade in the middle of a vertical scan. Protons elastically scattered
from hydrogen in the blades were detected by sets of forward and recoil
scintillator telescopes, the geometry of which was set to select only $\vec{p}p$
elastic events at 17.5$^{\circ}$ laboratory angle, near the maximum in the
parity allowed analyzing power, $A_y$.  During vertical scans the up-down
asymmetry was measured, and during horizontal scans the left-right asymmetry was
measured. Information from each blade was stored in 80 time bins of 80 $\mu$s
each, for a total time window of 6.4 ms per blade. As the radius of rotation is
215 mm, this corresponds to $\frac{6.4}{200}(2 \pi)(215)$ mm $= 43$ mm centered on
the nominal beam axis.  From the up-down or left-right asymmetry in each bin, a
distribution of transverse polarization across the beam profile was generated.
The two PPMs were operated with a 180$^\circ$ angular mismatch, producing 8
equally spaced profiles in 1/5 second, or 40 profiles per second. (This is also
the spin flip rate -- see Fig. \ref{timing})

The master clock for sequencing the whole experiment, including helicity
changes, was derived from 2500-line optical encoders mounted on the PPM drive
shafts. Each shaft encoder produced two 2500 cycle per revolution
square waves in phase quadrature, permitting the direction of rotation to be
determined from whether one signal was high or low at the positive going
transition of the other.  By recording both the rising and falling edges of
both quadrature signals, the PPM control electronics obtained an effective
resolution of 4 $\times$ 2500 = 10,000 lines.  The rotation of the two
PPMs was synchronized by electronic gearing using position information provided
by the shaft encoders. During steady state rotation at 5 revolutions per
second, the servo system was able to control the angle between the two PPMs to
$\pm 3$ milliradian, corresponding to one-third of the 1.6 mm target blade
thickness at the location of the proton beam.

\subsubsection{Liquid Hydrogen Target}

The $LH_2$ target had a flask 0.10 m in diameter and 0.40 m long. The target
scattered $\sim$4\% of the beam. It was important that the windows on the
hydrogen volume be flat and parallel to prevent motion of the proton beam from
creating intensity noise. The hydrogen was contained by 25 $\mu$m thick
stainless steel windows. Upstream and downstream of these inside the vacuum
vessel were two helium filled chambers, 21 mm thick on the upstream side and 10
mm thick on the downstream side, ensuring that the inner windows were flat. The
outer windows of the helium cells were 50 $\mu$m thick stainless steel. The
cryostat vacuum vessel was sealed with 25 $\mu$m thick copper windows. 

Although a thick target is desirable from the standpoint of increasing the
signal, it was also very important that the tails of the beam profile be well
contained in the 150 mm $\times$ 150 mm aperture of the detectors; beam blow-up
due to multiple Coulomb scattering limited the target flask length, as well as
the thickness of various entrance and exit windows and the thickness of the
upstream IPM foils. With the target full, the beam size increased from
approximately 4 mm ($\sigma$ of projected profile) at the center of the the
upstream detector to 15 mm at the center of the  downstream detector, and 22 mm
by its exit. 

Rapid circulation (5 $L/s$) of the $LH_2$ reduced density gradients and
prevented boiling.  A feedback loop using  fast and slow heaters controlled the
target temperature.  The fast heater responded to a beam current signal from
IPM1 (Sec. \ref{subsubsec:ipm}), and kept the heat load essentially constant
when the beam current changed. The slow heater made fine adjustments to hold
the LH$_2$ temperature at 19.3 K to within $\pm 0.2$ K over a several week data
taking period.

\subsubsection{Transverse Field Ionization Chambers}
\label{TRICs}

Each TRIC (a smaller earlier version is described in \cite{Stocki93})
consisted of a cylindrical enclosure filled with 750 liters of ultra-high
purity hydrogen gas at a pressure of about 150 Torr, and contained an upper
cathode plate operated at -8 kV, and a grounded lower signal plate. The TRICs
also contained field shaping electrodes plus guard rings to ensure a uniform
sense region, 150 mm wide by 150 mm high by 600 mm long between the parallel
electrodes. High-Z, low-energy spallation products from the chamber windows can
cause large fluctuations in the signal, so the entrance and exit windows were
located approximately 900 mm from the center to prevent spallation products
from entering the active region. Other considerations in the design of the
TRICs were noise due to delta ray ($\delta$-ray) production and ion pair
recombination. Recombination is reduced by lower gas pressure and higher
voltage, both of which reduce the space charge density in the active volume. In
practice, compromises must be made because too high a voltage causes increased
noise from corona discharge and low pressure reduces the desired signal.
$\delta$-rays are electrons produced by collisions of the protons with the
detector gas. The $\delta$-ray signal is noisy, and the major contribution to
this signal is from $\delta$-rays at large angles. The transverse dimension of
the ion chamber was a trade-off between a small transverse size which would
minimize $\delta$-ray noise and a large transverse dimension which accepts all
the beam. The 150 mm transverse dimension was the smallest that gave an
acceptably low sensitivity to beam size modulation  based on simulations.  The
main signal from the proton beam increases with the length of the active
region. This was limited by the available space in the beamline; 600 mm was the
longest practical length.

The noise in the ion chamber signals has two incoherent contributions -- shot
noise from the statistical nature of the proton beam, and noise contributed by
the chamber itself. The noise figure, $\alpha$, expresses the chamber noise as
a fraction of the shot noise. When the difference between the upstream and
downstream chambers is taken, most of the shot noise contribution disappears
because, except for the 4\% scattered by the target, each upstream proton also
passes through the downstream chamber. The chamber noise, on the other hand,
does not cancel, and the running time is dominated by chamber noise. In this
experiment the counting time was approximately 15 times that which would have
been expected from counting statistics alone. The fact that the run time was
dominated by chamber noise resulted in the seemingly paradoxical result that
better precision was obtained by lowering the beam current, because this
reduced the detector noise figure. In a series of test runs at progressively 
reduced beam currents, the $A_z$ distribution became narrower until
approximately 100 nA.  The 200 nA selected for running was a compromise between
lower current for better precision on $A_z$ and higher current for better
precision from the PPMs, which {\em were} limited by counting statistics.

\section{ Data Acquisition}

The data were taken in $\frac{1}{5}$ second (200 ms) cycles, each cycle
consisting of eight $\frac{1}{40}$ second (25 ms) spin states arranged in the
pattern ($+--+-++-$) or ($-++-+--+$). This pattern makes the result insensitive
to linear or quadratic drifts. The cycles were further arranged in an eight
cycle (64 spin state) ``supercycle'', with the starting state of each cycle
following  the same ($+--+-++-$) or ($-++-+--+$) pattern.  The initial spin
state of each supercycle was chosen at random. In the frequency domain, this
switching pattern contains odd multiples of 5 Hz, with the largest harmonic
content at 15 Hz. The data acquisition produced on-line values for the
amplitude and phase of the dominant 15 Hz component. This was derived from a 16
spin state cycle. The 64 spin state supercycle gave the option of using more
advanced digital filtering schemes on the difference signal, but these were
only used during early development runs.   20\% of the data were taken with all
the spin flipping equipment running, but with the pumping lasers blocked with a
shutter to guarantee zero polarization. In addition, as mentioned earlier, half
of these ``spin off'' data were taken with an artificially enhanced intensity
modulation synchronized with the spin flip.

\begin{figure}
\includegraphics[width=\linewidth]{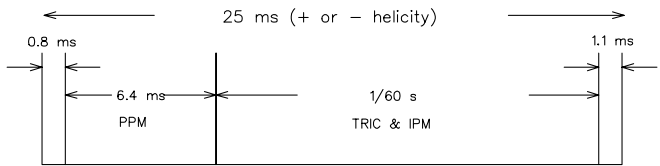}
\caption{The timing of detector readout intervals within one spin
state. The initial 0.8 ms allows for ion source settling time. The 6.4 ms
PPM interval corresponds to a 43 mm scan centered on the beam axis. The
TRIC interval was 1/60 second to eliminate sensitivity to 60 Hz or
harmonics of 60 Hz.}
\label{spinstate}
\end{figure}

As shown in Fig. \ref{spinstate}, each 25 ms spin state was divided into
polarization measuring and asymmetry measuring intervals. Following a short
dead time to allow the mechanical etalons which change the ion source spin
state time to stabilize, was a 6.4 ms window during which one of the 8  blades
of the scanning polarimeters passed through the beam.   The TRIC and IPM
signals were then integrated over exactly 1/60 second to eliminate sensitivity
to 60 Hz or harmonics of 60 Hz. The dead interval at the end of each state was
intended as a buffer to absorb any timing jitter due to imperfect rotation
speed of the PPMs. As it turned out, the timing jitter was less than 0.1 ms and
this buffer zone was more than adequate. 

The minimum data set for which a full set of helicity correlated beam
properties could be extracted was a 0.4 s ``event pair'' corresponding to two
full 360$^\circ$ rotations of the PPMs, as this gave both spin states for each
PPM blade.

As noted in the ion source section, the spin state was transmitted as a
frequency to prevent coupling the spin flip signal into the data acquisition.
The effectiveness of this isolation was checked by running the complete data
acquisition system including the ion source, but with detector signals supplied
by a battery. False $A_z$ from electronic crosstalk was found to be less than
$4 \times 10^{-9}$, an upper limit determined by the statistics of the test
(crosstalk was probably less).

The rotation speed of the PPMs is locked to a signal derived from the zero
crossing of the 60 Hz AC line. Since the 0.2 s taken for one PPM rotation is
exactly 12 cycles of the 60 Hz line, one would expect a given PPM blade to
always pass through the beam at the same phase of the AC line. To prevent
this, a small controlled phase slip was introduced.  The rate of slip was
programmable. For data taking it was set for one complete cycle of the 60 Hz
line in 18 minutes.  Although the phase of one PPM relative to the other is
``quantized'' by the finite resolution of the encoders, the angular frequency
of rotation is not, and in principle can be locked to a square wave of any
frequency. In practice, roundoff error in the motor control computer limited
the choices of the reference frequency. The 18 minutes for 360$^\circ$ of slip
corresponded to one of the acceptable values. 

In addition to the regular data taking runs, many dedicated control
measurements were made to measure the sensitivity of the apparatus to helicity
correlated beam properties, specifically: position, size, intensity, transverse
polarization, and energy modulation at the ion source.

\begin{figure}
\includegraphics[width=85mm]{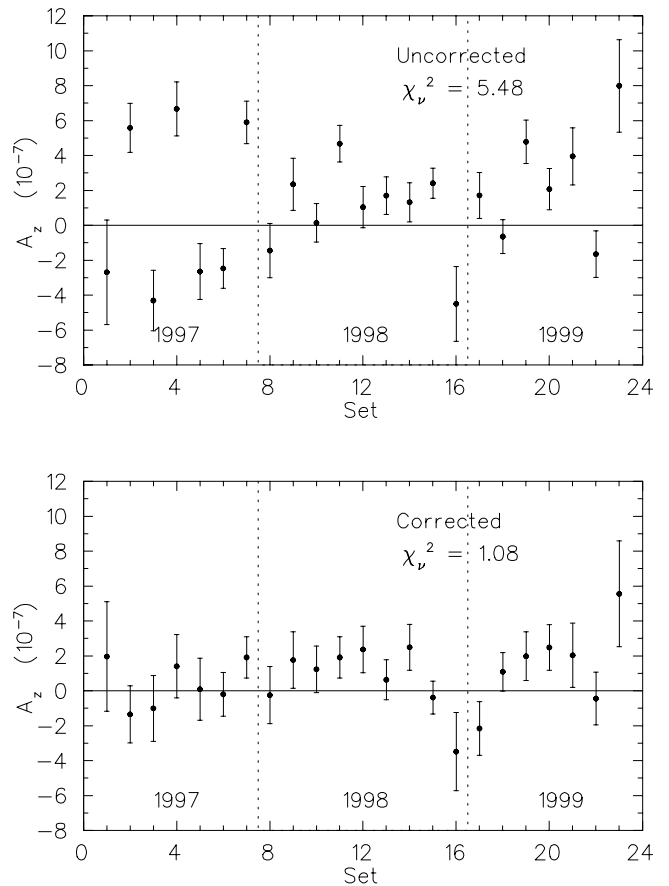}
\caption{E497 results before and after correction. The data are divided
into 23 sets of alternating beamline tune. The top panel shows the
results after beam quality cuts, but before the data were corrected for
systematic errors. The bottom panel shows the data after correction. The
error bars in the bottom panel are slightly larger due to uncertainties
in the corrections.}
\label{data}
\end{figure}

\section{ The Data}

\subsection{Raw Data Set}
The parity data used in the final analysis were acquired during
three major data taking periods -- February-March 1997, July-August
1998, and May-June 1999. The data were recorded as a series of
``runs'' of approximately one hour each. Statistics for the combined set
are

\begin{itemize}
\item 3.8 million event pairs
\item 375 runs in positive beamline helicity
\item 368 runs in negative beamline helicity
\item 80 position modulation runs
\item 81 size modulation runs
\item 40 energy modulation runs
\item 109 neutral axis scans
\end{itemize}

The runs were grouped into 23 sets, a new set being started following each
re-tune of the beamline (usually involving a reversal of the helicity at
the parity apparatus relative to the ion source).

\subsection{Reduced Data Set}

During data taking every effort was made to maintain the optimum beam
conditions (quiet beam, good longitudinal polarization, low transverse
polarization, low first moments of transverse polarization, beam on axis,
stable position and size, low intensity modulation), and to stop data
taking to correct the beam and cyclotron tune when these conditions were not
met. Nevertheless, some data were recorded for which conditions were not ideal.
The raw data sample was therefore subjected to data quality cuts prior to
inclusion in the final analysis.

The data quality cuts were conservatively chosen to ensure that corrections to
the reduced data set based on critical helicity-correlated beam parameters
would be consistent with sensitivities measured in ancillary calibration
experiments. The data quality cuts first eliminated 46 data runs for which the
TRIC difference signal was anomalously noisy, indicating unstable beam
conditions.  They also eliminated any data for which diagnostic monitor outputs
could be considered spurious. The cuts used are summarized in Table \ref{cuts}.
There were both ``hard cuts'' at a fixed value of beam parameter and ``soft
cuts'' at $\pm 3 \sigma$ from the mean value. The entire data reduction process
reduced the size of the total set by 30\%, but significantly improved the
quality of the data sample. The $\chi_\nu^2$ for the uncorrected data of the 23
sets went from 11.3 per degree of freedom before the cuts were made to 5.5
after the cuts. It is important to emphasize that the approach used to
determine the sensitivity to intrinsic polarization moments, which was the
dominant systematic error correction, required consistent and stable beam
conditions -- in particular, the ratio of beam sizes at IPM1/IPM2 and the ratio
of intrinsic polarization moments at PPM1/PPM2 must be constant over the data
sample used to determine the sensitivities, or the method is invalid.

It is important to emphasize that no data cuts were applied to the parity
violating asymmetry $A_z$ itself. The ultimate test of the consistency and
validity of the data analysis and corrections procedure is the quality of the
corrected data set shown in figure \ref{data}. The top panel in Fig. \ref{data}
shows the results for the reduced data set, before corrections were made for
systematic errors. The bottom panel shows the corrected results from which the
final $A_z$ was obtained. This corrected data set was subjected to a regression
analysis in which residual sensitivities to helicity correlated beam parameters
were explored. No statistically significant residual sensitivities were found,
lending confidence to the interpretation of the results. Details of the
corrections procedure are given in the following sections.

\begin{table}
\caption{Summary of data reduction cuts. Except for the beam
intensity, the center of the acceptance window for each parameter is
the measured centroid of that beam parameter's distribution. The hard
cuts are wide enough to include at least four standard deviations.}
\begin{tabular}{lr}
\toprule
 &  \\
Beam Parameter  & Acceptance Window \\
 &  \\
\colrule
Neutral axis $x,y$     &   $\pm 0.3$ mm         \\
Position               &   $\pm 3\sigma$ (soft) \\
Width                  &   $\pm 1.0$ mm         \\
Skew                   &   $\pm 0.2$ mm         \\
Intensity              &     196-204 nA         \\
Intrinsic Moments      &   $\pm 3.0$ mm         \\
\botrule
\end{tabular}
\label{cuts}
\end{table}

\section{Sources of Error}
Sources of error can be divided into those that are related to the beam
helicity and those that are not. Beam property changes that are synchronized
with spin flip (``coherent'' or ``helicity correlated'' changes) can shift the
centroid of the $A_z$ distribution and cause a false signal of parity
violation. Changes that are not helicity correlated, such as detector noise and
random variation in beam properties, do not bias the result, but they increase
the run time required to reach a given precision.

\subsection{Random Changes}

The ultimate limit to the statistical precision of the experiment was set by
the counting statistics of the scattered protons.  For a target scattering 4\%
of the 200 nA beam, the rate of scattered protons is 50 GHz, which would make
it possible to measure $A_z$ to $\pm 0.2 \times 10^{-7}$ in 20 hours if
individual scattered protons could be counted. However, 50 GHz was too high for
direct counting, and so current mode detection was used. As noted already in
the ion chamber section, the counting time was dominated by detector noise.
Other random variations in beam properties such as intensity and position also
contributed noise and further increased the required run time. It proved to be
a net advantage to devote significant beam development time to producing quiet,
stable, beam.

\subsection{Helicity Correlated Changes}

The approach to minimizing and correcting for the false $A_z$ signal due to
helicity correlated beam property changes was as follows:
\begin{itemize}
\item Careful design and operation of the TRIUMF optically pumped polarized ion
source and cyclotron made it possible to change the spin direction with very
little effect on the other beam properties.
\item The design of the parity equipment and the operating conditions of the
experiment were carefully chosen to minimize the sensitivity to helicity
correlated changes.
\item Calibration runs determined the sensitivity to helicity correlated
modulations.
\item The beam properties were continuously monitored during data taking so the
actual helicity correlated changes were known, and appropriate corrections were
applied.
 \end{itemize}

\section{ Correction for Systematic Errors}

The effects of modulations in beam intensity, position, size,
transverse polarization, and energy were considered. 

Measurements made early in the experimental program revealed significant
correlations between beam parameters.  For example, horizontal and vertical
beam motion are often highly correlated. To be sure of extracting the correct
sensitivities, separate control measurements were made in which each beam
property in turn was artificially modulated, and the effect on apparent $A_z$
was recorded. The sensitivities obtained by this method were consistent with
sensitivities extracted by multilinear regression using the natural variation
of beam parameters. Most importantly, the regression analysis showed no
significant sensitivity to {\em products} of beam modulations, so the false
$A_z$ was the sum of the contributions of the individual coherent modulations. 

For the sensitivity calibrations made by varying one parameter at a time,
measurements were made using modulations of different size and sign. The false
asymmetry was linear in the modulation, passing  through zero at zero
modulation. To determine the sensitivities used for data reduction, large
values of modulation were used to give accurate results in a short time. To
determine the sensitivity to first moments of transverse polarization, which
gave the largest false asymmetry, much larger transverse polarizations were
used for calibration than were present in the parity beam. This was justifiable
because false asymmetries must be linear in  polarization, and the use of large
polarizations allowed the displacements needed for the calibration to be
reduced. The measured false asymmetry was linear in polarization moment. 

\subsection{Coherent intensity modulation (CIM)}

As the energy of the beam was 27 MeV lower at TRIC2 than at TRIC1, the TRIC2
gas pressure was lowered relative to TRIC1 to equalize ion chamber currents.  A
hardware gain on the TRIC1 signal was then adjusted to minimize common-mode
noise in the amplified difference signal, $1000\times(I_{2} - gI_{1})$\, which
was digitized, where $g$\ is the hardware gain; $g$ is very close to unity. 
The helicity correlated part of the difference, which should be proportional to
$A_z$, nevertheless contains a coherent current modulation component due to
imperfect common mode rejection arising from gain mismatch and nonlinearity of
the ion chambers. If the current signal from TRIC2 is $I_{2}^{\pm} = a
I_{1}^{\pm}(1 - S(1 \pm P_{z}A_{z}))$ then, expressing the difference in
current for the two helicity states as $I_{1}^{\pm} = I_{1} \pm \delta I_{1}$,
one can define an ``analog asymmetry'',

\begin{eqnarray}
\epsilon_{a} &=&
\frac{(I_{2}^{+} - gI_{1}^{+}) - (I_{2}^{-} - gI_{1}^{-})}{2I_{1}}
\nonumber\\
&=& -S a P_{z}A_{z} + \left(a T - g +
I_{1}T\frac{da}{dI_{1}}\right) \frac{\delta I_{1}}{I_{1 }},
\end{eqnarray}

\noindent where $S$ is the nuclear scattering probability in the LH$_{2}$\
target ($S \simeq 0.04$) and $T = 1 - S$.  The quantity $a$ is a function of
TRIC1 and TRIC2 gas gains, and is nominally equal to $1/T$; it was adjusted
with TRIC2 gas pressure.  The hardware gain $g$ was set by zeroing the false
parity signal when the beam current was modulated by the photodetachment laser.
Because of the nonlinearity term, $g$ must be reset if the beam current
changes. 

\begin{figure}
\includegraphics[width=85mm]{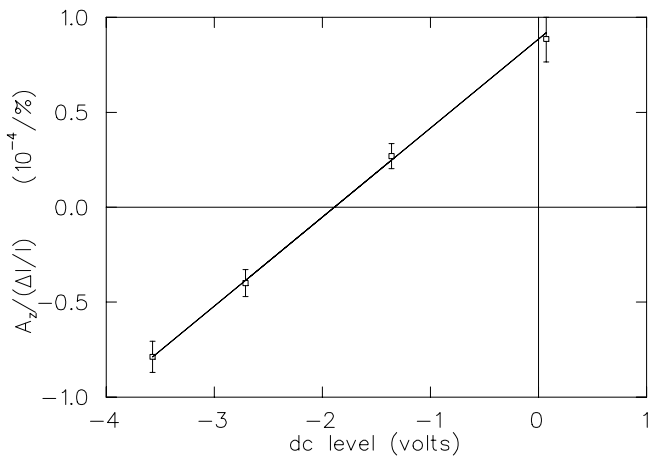}
\caption{A scan of the subtractor box settings taken during the July 1998
running period. The plot shows the sensitivity to CIM as a function of the
subtractor box DC level. The DC level is a sensitive indication of the
setting of the fine gain control; a fractional gain change of $10^{-4}$
causes a shift of 0.7 V in the DC level. This particular scan gave an ideal
DC level of -1.9 V for zero sensitivity to CIM.}
\label{cimtune}
\end{figure}

A beam current stabilization system was implemented which restricted beam
current excursions to the range $200 \pm 2$ nA \cite{Zelen98}. The current
stabilization system took a current signal from the upstream ionization chamber
(TRIC1) and fed a correction voltage to an electrostatic quadrupole lens just
upstream of a set of slits in the injection line. The injection line was tuned
with a slight excess of current so that, at the set point, about 10\% of the
beam was skimmed by the slits. To exclude coupling of current modulation to
position and energy modulations, the sampling rate was slow (0.5 Hz), much
slower than the spin-flip rate of 40 Hz. The active range of the current
stabilization loop was quite low ($\pm10$\%) to avoid correcting drifts caused
by large excursions of cyclotron tune. The current loop operation was a
sensitive indicator of cyclotron stability, and operators made adjustments to
the cyclotron tune when necessary. Even under conditions of constant beam
current, periodic adjustments of $g$ were needed to compensate for small drifts
in the ion chamber gains. The setting for best common mode rejection was
checked periodically by turning on the photodetachment laser, creating a ($\sim
0.1$\%) coherent intensity modulation, and performing a scan of subtractor box
settings. The results of such a calibration are shown in Fig. \ref{cimtune}.
Because the difference signal was multiplied by 1000, the DC level was a
sensitive indication of the setting of the hardware fine gain. A fractional
change of $10^{-4}$ in the hardware fine gain caused a DC shift of 0.7 V in the
subtractor box output.

\begin{figure}
\includegraphics[width=85mm]{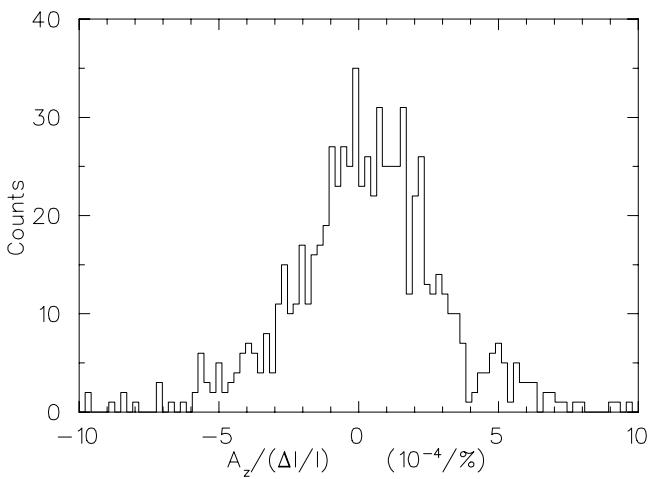}
\caption{CIM data for a typical one hour run. The plot shows a histogram of
the sensitivity to $\frac{\delta I}{I}$. Data have been excluded for
which $\left|\frac{\delta I}{I}\right|<0.06$\%. The mean is 0.23 and the
standard deviation 2.6, in units of ($10^{-4}$/\%).}
\label{cimdata}
\end{figure}

During data taking, coherent intensity modulation (CIM) was measured
continuously by the ion chamber (TRIC1) upstream of the target. As noted in
section \ref{sec:oppis}, this was  normally less than a few parts in $10^5$.
Because the sensitivity to CIM changed between the dedicated CIM runs used to
re-set the subtractor box, the sensitivity to CIM, like the CIM itself, was
measured during data taking. Periods of enhanced (~0.1\%) coherent intensity
modulation were interleaved with the main data. A typical sensitivity was
$A^{false}_z = (1.6 \times 10^{-4})(I_b - 200)(\frac{\delta I}{I})$ where $I_b$
is the beam current in $nA$. Table \ref{cimsens} shows the average
sensitivities to coherent intensity modulation for the different running
periods, and Fig. \ref{cimdata} shows a histogram of the sensitivities for a
typical ($\sim 1$ hour) run.

\begin{table}
\caption{Average sensitivities to Coherent Intensity Modulation. The
sensitivities are determined from all CIM data with
$\vert \frac{\Delta I}{I} \vert \geq 0.06$ \%.
}
\begin{tabular}{lcr}
\toprule
 Set   & Sensitivity $(10^{-7}/$\%) & CIM events \\
\colrule
 1997   & $36 \pm 19$ &  83,636  \\

 1998  & $138 \pm 8$ & 106,520  \\

 1999 & $73 \pm 11$ &  90,955  \\

\botrule
\end{tabular}
\label{cimsens}
\end{table}

\subsection{Coherent position modulation}
Coherent position modulation was measured by two of the three IPMs. During
the 1997 running period, IPM1 and IPM3 were used because their separation
was greater and, all else being equal, position control should have been
better. Unfortunately IPM3 picked up noise from the liquid hydrogen target
circulation fan, so for the 1998 and 1999 data, IPM1 and IPM2, located 1.8
m apart along the beamline, were used. Using two IPMs permits measuring
both tilts and parallel shifts of the beam. The measured coherent
modulations are shown in Tables \ref{sizpos97} and \ref{sizpos9899}. The
sensitivity to beam motion was measured in separate control measurements
during which the beam position was modulated in a variety of ways using the
FCSMs.

\begin{figure}
\includegraphics[width=85mm]{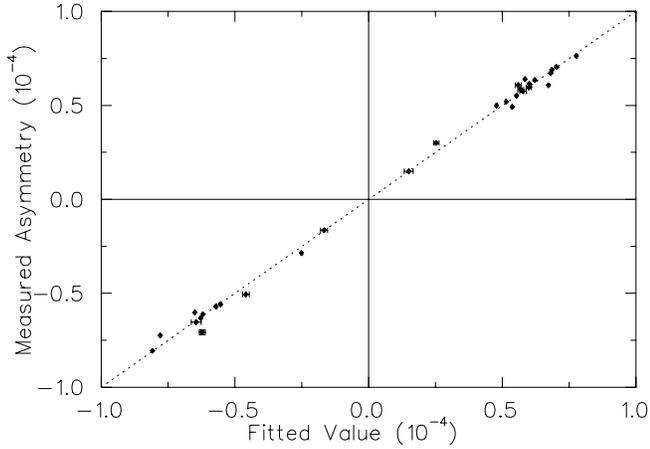}
\caption{Calibration of sensitivity to coherent position modulation.
The abscissa shows the false asymmetry based on
sensitivities extracted from calibration runs and the ordinate
shows the asymmetry actually measured.}
\label{posmod}
\end{figure}

The false analyzing power arising from helicity correlated beam
position was parameterized as
\begin{eqnarray}
\Delta A_z =
  \left( \frac{\Delta x_1 + \Delta x_2}{2} \right) (a_x x_1+b_x x_2+c_x)
\nonumber\\
+ \left( \frac{\Delta x_1 - \Delta x_2}{2} \right) (d_x x_1+e_x x_2+f_x)
\nonumber\\
+ \left( \frac {\Delta y_1 + \Delta y_2}{2} \right) (a_y y_1+b_y y_2+c_y)
\nonumber\\
+ \left( \frac {\Delta y_1 - \Delta y_2}{2} \right) (d_y y_1+e_y y_2+f_y)
\end{eqnarray}
where $\Delta x$ and $\Delta y$ are the horizontal and vertical helicity
correlated beam motion, $x$ and $y$ are the beam position and the 1 and
2 subscripts refer to IPM1 and IPM2. The parameters $a$ to $f$ were
extracted from a fit to the calibration data. Details may be found in
Ref. \cite{Bland01}. Figure \ref{posmod} shows the results of such a fit.
The abscissa shows the false asymmetry predicted using Eq. 2 with the
fitted parameters, and the ordinate shows the false asymmetry actually
measured with the calibration data.

\subsection{Coherent beam size modulation}
Coherent beam size modulation was also measured by the two IPMs, and the
observed values are listed in Tables \ref{sizpos97} and \ref{sizpos9899}.
The sensitivity to beam size was determined by control measurements in
which the beam size was intentionally modulated by driving the FCSMs as
quadrupoles.

\begin{figure}
\includegraphics[width=85mm]{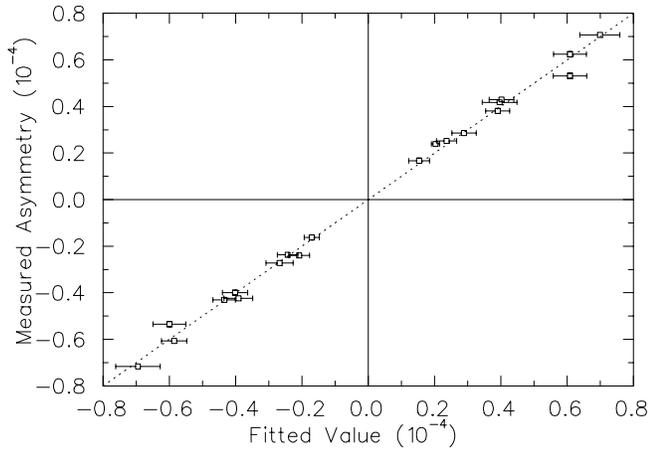}
\caption{Calibration of sensitivity to coherent size modulation. The
abscissa shows the false asymmetry based on
sensitivities extracted from calibration runs and the ordinate
shows the asymmetry actually measured.}
\label{sizemod}
\end{figure}

The false analyzing power due to beam size modulation was expressed as

\begin{eqnarray}
\Delta A_z = \alpha_x \sigma_{x_1} \Delta \sigma_{x_1}
+ \beta_x \sigma_{x_2} \Delta \sigma_{x_2}
\nonumber\\
+ \alpha_y \sigma_{y_1} \Delta \sigma_{y_1}
+ \beta_y \sigma_{y_2} \Delta \sigma_{y_2}
\end{eqnarray}

\noindent where $\sigma$ is the RMS beam size and $\Delta \sigma$ is the
helicity correlated change in beam size. As with position modulation, the
parameters were extracted from a fit to the calibration data.  Because the
quadrupole configuration of the FCSMs caused the beam to be steered somewhat
when the size was modulated, correction for position modulation had to be
included in the fit. Figure \ref{sizemod} shows the results of a fit to a
series of size modulation calibration runs.

\begin{table}
\caption{Summary of Position and Size Modulation for the 1997 data
set. These data used IPM3 for fast position control. Changing the
Wien filter sets the spin direction in the cyclotron relative to the
ion source and the beamline sets the helicity at the parity apparatus
relative to the cyclotron.}
\begin{tabular}{lcccr}
\toprule  & \multicolumn{2}{c}{Wien filter $+$} &
\multicolumn{2}{c}{Wien filter $-$} \\
\colrule
 Beam Line:                     &  $+$                &  $-$
       &  $+$                   &    $-$           \\
\colrule
 $\Delta \sigma_{x_1} (nm)$     & $119 \pm 49$         & $-137 \pm 43$
       & $-101 \pm 59$       & $8 \pm 48$          \\
 $\Delta \sigma_{x_3} (nm)$     & $65 \pm 79$          & $-156 \pm 76$
       & $-110 \pm 118$      & $-45 \pm 91$        \\
 $\Delta \sigma_{y_1} (nm)$     & $126 \pm 44$         & $-212 \pm 44$
       & $-36 \pm 54$        & $41 \pm 45$         \\
 $\Delta \sigma_{y_3} (nm)$     & $81 \pm 75$          & $-200 \pm 72$
       & $-54 \pm 116$       & $-11 \pm 84$        \\
 $\Delta x_1 (nm)$              & $5 \pm 33$           & $19 \pm 28$
         & $-34 \pm 40$        & $38 \pm 36$       \\
 $\Delta x_3 (nm)$              & $-30 \pm 34$         & $-40 \pm 34$
        & $10 \pm 46$         & $4 \pm 37$        \\
 $\Delta y_1 (nm)$              & $-46 \pm 26$         & $89 \pm 27$
         & $-7 \pm 33$         & $-10 \pm 27$      \\
 $\Delta y_3 (nm)$              & $-27 \pm 36$         & $-12 \pm 33$
        & $24 \pm 55$         & $-78 \pm 43$      \\
\botrule
\end{tabular}
\label{sizpos97}
\end{table}

\begin{table}
\caption{Summary of Position and Size modulation for the 1998 and 1999 data
sets. IPM2 was used for fast position feedback, replacing IPM3. The Wien filter
was $+$ for all these data, producing spin-up in the cyclotron. The beamline
rotates this to $+$ or $-$ helicity at the parity apparatus.}
\begin{tabular}{lcr}
\toprule
 Beam Line:                   &  $+$             &        $-$ \\
\colrule
 $\Delta \sigma_{x_1} (nm)$   & $-17 \pm 38$     & $0 \pm 35$ \\
 $\Delta \sigma_{x_2} (nm)$   & $8 \pm 8$        & $-35 \pm 7$ \\
 $\Delta \sigma_{y_1} (nm)$   & $-6 \pm 37$      & $1 \pm 35$ \\
 $\Delta \sigma_{y_2} (nm)$   & $17 \pm 7$       & $-37 \pm 5$ \\
 $\Delta x_1 (nm)$            & $1 \pm 14$       & $-1 \pm 6$ \\
 $\Delta x_2 (nm)$            & $7 \pm 11$       & $-9 \pm 5$ \\
 $\Delta y_1 (nm)$            & $17 \pm 11$      & $3 \pm 8$ \\
 $\Delta y_2 (nm)$            & $16 \pm 10$      & $2 \pm 7$ \\
\botrule
\end{tabular}
\label{sizpos9899}
\end{table}

\subsection{Transverse polarization}

\begin{figure}
\includegraphics[width=85mm]{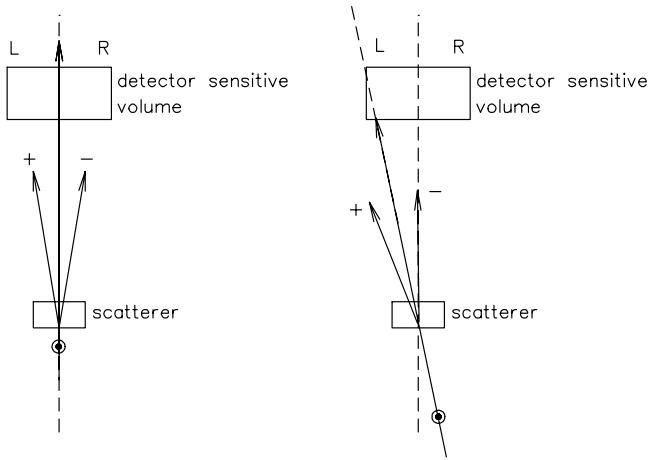}

\caption{The effect of transverse components of polarization. The combination
of a transverse polarization component with a beam displacement from the
neutral axis will cause a false signal of parity violation.}

\label{firstmom}
\end{figure}

If the proton spin is not perfectly longitudinal, the small transverse
component will reverse with helicity. This can couple to the relatively large
($\sim 0.3$) parity allowed analyzing power to cause a false parity violating
signal. Figure \ref{firstmom} uses a pencil beam with a vertical transverse
polarization to illustrate the mechanism. Because the parity allowed analyzing
power is positive, slightly more beam will be scattered to the left in the
positive helicity state, and to the right in the negative helicity state. If
the beam passes through the center (neutral axis) of the detector, as shown in
the left-hand panel, the response will be the same for both helicity states and
no false effect arises. If, on the other hand, as shown in the right-hand
panel, the beam does not pass through the center of the detector, more signal
will be recorded in one helicity state than in the other. To a very good
approximation, the effect is found to be proportional to the size of the
transverse component multiplied by the distance the beam is off center at the
detector -- i.e. to the {\em first moment} $\langle x \rangle \langle P_y
\rangle$ of transverse polarization at the detector. 

In a field-free region, a real particle beam of finite extent is made up of a
bundle of straight rays like the pencil beam in this example. The first moment
for the beam is the linear sum of the first moments of the individual rays. 
Since the first moment for a ray is proportional to the distance of the ray
from the zero axis and varies linearly with distance along the beam, changing
sign where the ray crosses zero, the first moment for a particle beam will vary
linearly with distance along the beam. A real beam can have a substantial first
moment of transverse polarization even if the net transverse polarization is
zero. In the example of Fig. \ref{firstmom}, the vertical polarization could be
``up'' on the right side of the beam and ``down'' on the left side.

\begin{figure}
\includegraphics[width=85mm]{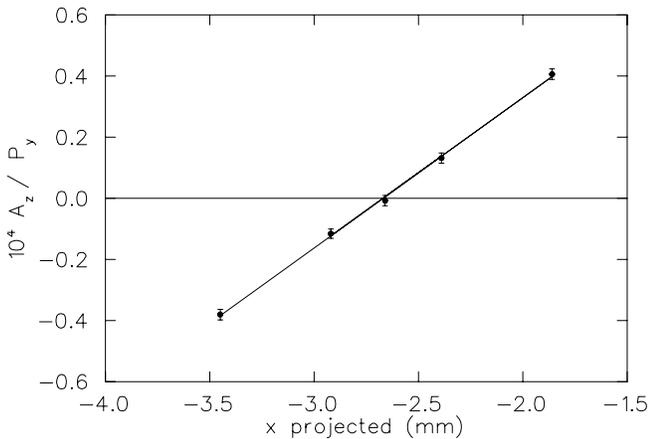}
\caption{A horizontal neutral axis scan. A vertically polarized beam ($P_y
= 0.76$) was scanned horizontally and the sensitivity to $P_y$ plotted as a
function of $x$ projected to the ``magic $z$'' ($z$ location where the
false $A_z$ does not depend on angle). A beam with $x_{proj}=-2.7$ mm was
found to produce no false signal from vertical polarization components.}
\label{pycal}
\end{figure}

To determine the first moment sensitivities, test runs were made with pure
vertical and pure horizontal polarization.  By scanning the vertically
polarized beam horizontally it was possible to generate known moments, 
$\langle x \rangle \langle P_y \rangle$, and, by scanning the horizontally
polarized beam vertically, known  $\langle y \rangle \langle P_x \rangle$
moments could be generated. Although in principle the first moments must be
known at the two detectors, because first moments vary linearly with distance
along the beamline, knowledge of the first moments at {\em any} two locations
is sufficient.

The transverse polarization profiles were measured by the two Polarization
Profile Monitors (PPMs in Fig. \ref{general}) located 1.8 m apart along the
beamline and, for convenience, the first moment sensitivities were expressed in
terms of the sensitivities at the PPMs. The first moment scans also defined the
beam trajectory for which there was zero sensitivity to average transverse
polarization. Such a trajectory was referred to as a {\em polarization neutral
axis}. Polarization neutral axes were determined in both the horizontal and
vertical directions and, during data taking, the beam was held on this neutral
axis by the servo system. Figure \ref{pycal} shows an example of a horizontal
neutral axis scan. Another consequence of the linear behavior of the first
moments with distance, $z$, is the presence of a ``magic $z$'' downstream of
TRIC2. {\em Any} beam with zero first moment at this $z$ will cause no false
effects. Also notice that the measurements shown in Fig. \ref{pycal} confirm
the linear relation between first moment and false $A_z$.

In analyzing the effect of first moments, {\em extrinsic} first moments,
$\langle x \rangle \langle P_y \rangle$ and $\langle y \rangle \langle P_x
\rangle$, caused by a beam which has some net transverse polarization and whose
centroid is displaced from the neutral axis, were treated separately from {\em
intrinsic} first moments, $\langle xP_y \rangle$ and $\langle yP_x \rangle$,
which do not depend on the position of the beam, but rather arise from the
distribution of components within the beam. By holding the beam on the
polarization neutral axis it was possible to virtually eliminate corrections
for {\em extrinsic} first moments.  {\em Intrinsic} first moments, on the other
hand, are independent of beam position, arise in the cyclotron and beamline,
and were very hard to control.

The false asymmetry arising from extrinsic and intrinsic first moments
can be written 

\begin{eqnarray}
\Delta A_z &=
\langle P_x \rangle
 [a_1(\langle y_1 \rangle - y_1^{na})
&+ a_2(\langle y_2 \rangle - y_2^{na})]
\nonumber \\
&+\langle P_y \rangle
 [b_1(\langle x_1 \rangle - x_1^{na})
&+ b_2(\langle x_2 \rangle - x_2^{na})]
\nonumber \\
&+ a_1 \langle y P_x \rangle _1 &+ a_2 \langle y P_x \rangle _2
\nonumber \\
&+ b_1 \langle x P_y \rangle _1 &+ b_2 \langle x P_y \rangle _2
\end{eqnarray}

\noindent where the first two lines are the contribution from {\em
extrinsic} first moments, and the second two lines are the contribution
from {\em intrinsic} first moments. Note that the first moment
sensitivities, $a_1$, $a_2$, $b_1$, and $b_2$, are identical for intrinsic
and extrinsic moments.

The results of the neutral axis scans are shown in Table \ref{fmsens}. The
quantities $(-a_1 y^{na}_1 - a_2 y^{na}_2)$ and $(-b_1x^{na}_1 - b_2
x^{na}_2)$ depend on what axis $x$ and $y$ are measured from. They
are zero if $x_1$, $y_1$, $x_2$, and $y_2$ are measured relative to the
neutral axis. One notes that the sensitivities to first moments are very
consistent from run period to run period. In units of $10^{-7}$ per
$\mu m$, the sensitivity to $\langle y \rangle \langle P_x \rangle$ is 1.8
at PPM1 and -2.5 at PPM2. The sensitivity to $\langle x \rangle \langle
P_y \rangle$ is -1.5 at PPM1 and 2.0 at PPM2.

\begin{table}
\caption{Fitted first moment sensitivities from the neutral axis scans.}
\begin{tabular}{lccr}
\toprule
 & & \\
Sensitivity (10$^{-7}$)	& Feb97 & Jul98 & May99 \\
\colrule
$a_1 (\mu m^{-1}$) & $1.91 \pm 0.03$  & $1.85 \pm 0.01$  & $1.83 \pm 0.02$ \\
$a_2 (\mu m^{-1}$) & $-2.47 \pm 0.01$  & $-2.47 \pm 0.01$  & $-2.44 \pm 0.01$ \\
$(-a_1 y^{na}_1 - a_2 y^{na}_2)$
& $-0.55 \pm 0.01$    & $-1.15 \pm 0.01$  & $-0.26 \pm 0.01$ \\
$b_1 (\mu m^{-1}$) & $-1.51 \pm 0.03$ & $-1.49 \pm 0.01$ & $-1.49 \pm 0.02$ \\
$b_2 (\mu m^{-1}$) & $2.02 \pm 0.04$  & $2.00 \pm 0.01$   & $1.98 \pm 0.01$ \\
$(-b_1x^{na}_1 - b_2 x^{na}_2)$
& $1.31 \pm 0.03$   & $-0.35 \pm 0.01$    & $0.50 \pm 0.01$ \\ \hline \hline
 $\chi^2_{\nu}$ & 1.17  & 1.96 & 2.05 \\
\botrule
\end{tabular}
\label{fmsens}
\end{table}

There were no magnetic elements after the first PPM, so the first moments of
transverse polarization varied linearly with position along the beamline.
Furthermore, for a fixed setting of the upstream beamline magnets, the first
moments at PPM1 and PPM2 tended to scale together so that, over a wide range of
first moments, the {\em ratio} of first moments at PPM1 and PPM2 had a constant
value. Since the neutral axis scans showed that the first moment sensitivity at
PPM1 was of opposite sign to, and 75\% of the magnitude of, that at PPM2, if
the PPM2 first moment was always 75\% of the PPM1 first moment, then the {\em
effective} sensitivity to first moments would be zero. This would correspond to
a beam whose first moment of transverse polarization goes through zero at the
magic $z$. By adjusting the beam convergence, an attempt was made to achieve
this magic moment ratio between PPM1 and PPM2.  Unfortunately, the ability of
the PPMs to measure this ratio to sufficient precision in a reasonable amount
of time was very limited, so the cancellation was not perfect and the residual,
effective, first moment sensitivity had to be extracted from the data. This
could be done either by regressing the first moment ratio from the data and
using the sensitivities measured in the neutral axis scans, or by regressing
the correlation of $A_z$ with average first moment directly from the data. The
two methods agreed with each other, but the latter method produced the smallest
statistical spread and was the method used to correct the data.

\subsection{Energy Modulation} 
\label{subsec:emod}
Since the beam energy in the downstream TRIC was on average 27 MeV lower
than in the upstream TRIC due to energy loss in the target,
helicity-correlated energy modulation caused a systematic error due to the
nonlinear energy dependence of the proton beam energy loss in the hydrogen
gas of the TRICs. The sensitivity to coherent energy modulations was
determined using an RF accelerating cavity placed upstream of IPM1 in the
beam line. The measured sensitivity of false $A_z$,  $(2.9 \pm 0.3) \times
10^{-8} eV^{-1}$,  was in excellent agreement with predictions based on the
variation of stopping power with energy.

Energy modulation of the extracted beam was caused by position modulations of
the radial intensity distribution at the stripping foil; this converted radial
position modulation of the injected beam to energy modulation of the extracted
beam. The primary coherent energy modulation produced in the source was
converted to position modulation in the injection beamline and then back to
energy modulation at the extraction foil. Direct measurements using a magnetic
spectrometer (1.2 m dispersion) in another beamline (4B) at TRIUMF showed the
energy modulation of the extracted beam to be approximately 100 to 200 times
greater than the energy modulation at OPPIS. During the parity runs this direct
measure could not be made, but frequent measures of energy modulation at OPPIS
and of $dA_z/dE_{oppis}$ were made. The $dA_z/dE_{oppis}$ sensitivity was
measured by applying a square wave voltage of 0.5 V amplitude to the
electrically isolated sodium ionizer in OPPIS. These $dA_z/dE_{oppis}$ runs
were then corrected for all known systematic errors, (dominated by $dI/I$), and
it was assumed that the residual false $A_z$ arose from energy modulation of
the extracted beam. A comparison of the measured $dA_z/dE_{oppis}$ to the
$dA_z/dE$ measured in the beamline with the RF cavity (Sec. \ref{subsec:emod})
indicated that the cyclotron amplified $dE_{oppis}$ by a factor of about 130,
in agreement with the magnetic spectrometer measurements. 

The primary energy modulation caused by the optical pumping lasers was measured
by using an electrostatic beam energy analyzer in the polarized source and an
intensity profile monitor with 16 collector strips 2.5 mm wide, and with 3.0 mm
spacing to measure beam position modulation downstream of the steering
analyzing plates. The monitor was mounted on a remotely controlled swinging
arm. Two measurements of coherent position modulation for the right and left
monitor positions allowed separation of the energy and position modulation
components of the OPPIS beam. An accuracy of 0.2 meV could be achieved in ten
minutes of integration time. The modulation magnitudes were quite sensitive to
the pumping laser asymmetry between the two polarization states; after careful
laser tuning, the coherent energy modulation was reduced to 1-2 meV and the
coherent position modulation to the 20 nm level.

\begin{table}
\caption{Settings of the Wien filter and beamline for the 23 data sets.}
\begin{tabular}{lcr}
\toprule
 \multicolumn{3}{c}{set,(Wien filter, beamline)} \\
\colrule
   $1,(+,+)$ ~~     &  $ 9,(+,-)$ ~~    &  $17,(+,+)$    \\
   $2,(+,-)$ ~~     &  $10,(+,+)$ ~~    &  $18,(+,-)$    \\
   $3,(+,+)$ ~~     &  $11,(+,-)$ ~~    &  $19,(+,+)$    \\
   $4,(+,-)$ ~~     &  $12,(+,+)$ ~~    &  $20,(+,-)$    \\
   $5,(+,+)$ ~~     &  $13,(+,-)$ ~~    &  $21,(+,+)$    \\
   $6,(-,+)$ ~~     &  $14,(+,+)$ ~~    &  $22,(+,-)$    \\
   $7,(-,-)$ ~~     &  $15,(+,-)$ ~~    &  $23,(+,+)$    \\
   $8,(+,+)$ ~~     &  $16,(+,+)$ ~~    &                \\
\botrule
\end{tabular}
\label{helicities}
\end{table}

Although the frequent measurement of $\Delta E$ and $dA_z/dE_{oppis}$ helped to
set limits on the false $A_z$ from energy modulation of the extracted proton
beam, it is significant that the energy modulation of the extracted proton beam
could not be measured directly at the parity apparatus.  To cancel its effects,
use was made of the fact that when the beamline helicity is reversed, the
effects of true $A_z$ reverse, but the effects of energy modulation do not. 
This is because energy modulation can only arise in the ion source and
cyclotron. The magnets used to rotate the spin from ``up'' to positive helicity
or ``up'' to negative helicity are downstream of all sources of energy
modulation and do not affect the beam energy.  Data were generally taken with
alternating beamline helicity, and during the 1997 run data were also taken
with the Wien filter reversed, an independent method of reversing the proton
beam helicity relative to the ion source. The Wien filter and beamline settings
are detailed in Table \ref{helicities}. That the $\chi_{\nu}^2$ for the 23 sets
is only 1.08 following corrections, shows that the effects of uncorrected
systematic errors, including energy modulation, were small. Care was taken to
balance the amount of data taken in the two beamline helicities so that if some
false $A_z$ from energy modulation was present, it would tend to cancel in the
final weighted average. Information on the energy modulation at the ion source
and the sensitivity to this modulation was used to include the effects of
uncorrected energy modulation in the error budget, as described in more detail
in the next section.

\begin{table}
\caption{Overall corrections for systematic errors. The table shows
the average value of each coherent modulation, the net correction made for
this modulation, and the uncertainty resulting from applying the
correction.}
\begin{tabular}{lcr}
\toprule
 & & \\
 Property & Average Value & $10^7 \Delta A_z$  \\
          &               &  (correction) \\
 & & \\
\colrule
 $A_z^{uncorrected} (10^{-7})$ & ~~~~~~~$1.68 \pm 0.29(stat.)$ & \\
 $y*P_x (\mu m)$ & $-0.1 \pm 0.0$ & $-0.01 \pm 0.01$ \\
 $x*P_y (\mu m)$ & $-0.1 \pm 0.0$ & $0.01 \pm 0.03$ \\
 $\langle yP_x \rangle (\mu m)$ & $1.1 \pm 0.4$ & $0.11 \pm 0.01$ \\
 $\langle xP_y \rangle (\mu m)$ & $-2.1 \pm 0.4$ & $0.54 \pm 0.06$ \\
 $\Delta I/I (ppm)$ & $15 \pm 1$ & $0.19 \pm 0.02$ \\
 $position + size$ &           & $     0  \pm 0.10$ \\
 $\Delta E(meV at\, OPPIS)$&   7--15      & $  0.0  \pm 0.12$ \\
 electronic crosstalk &     & $ 0.0 \pm 0.04$ \\
 Total & & $0.84 \pm 0.17 (syst.)$ \\
\hline\hline
 $A_z^{corr} (10^{-7})$ &
\multicolumn{2}{c|}{$0.84 \pm 0.29(stat.) \pm 0.17(syst.) $} \\
 $\chi_{\nu}^2 (23 sets)$ &
\multicolumn{2}{c|}{1.08} \\
\botrule
\end{tabular}
\label{corr}
\end{table}
\section{Method of Applying The Corrections}

The false $A_z$ arising from a given coherent modulation was found by
multiplying the measured modulation by the sensitivity to that modulation.
Uncertainties from the variance of the corrected $A_z$ distribution are
referred to as ``statistical''. The uncertainty quoted is the standard
error in the mean of the $A_z$ distribution. Uncertainties in $A_z$
resulting from uncertainties in the sensitivities are referred to as
``systematic'' because an incorrect sensitivity will cause a systematic
shift in the mean of the $A_z$ distribution. These errors are, however,
statistical in nature, as they arose from statistical uncertainties in the
knowledge of the sensitivities to the various coherent modulations.

Table \ref{corr} summarizes the overall corrections to the parity data. To
produce the 23 corrected $A_z$ distributions shown in Fig. \ref{data}, the
following procedure was followed.

\begin{enumerate}

\item The data in each set were grouped into bundles of 10000 event
pairs per bundle.

\item Each bundle was corrected according to the observed coherent modulations
(except position and size) for that bundle, giving a corrected $A_z$ for each
bundle. The variance of corrected $A_z$ values in a set determined an error bar
for that set. This is reported as the ``statistical'' uncertainty. No
corrections were made for position and size, because the net correction was
consistent with zero and when the corrections were applied it was found that
they slightly {\em increased} the residual correlations of $A_z$ with position
and size, as well as increasing the variance of the corrected $A_z$
distribution.

\item The uncertainties in the nil correction for position and size modulation
were added in quadrature to each of the 23 sets. This is included in what is
reported in Table \ref{corr} as ``systematic'' uncertainty.

\item The uncertainties resulting from uncertainties in the various
sensitivities were added in quadrature to the corresponding data. The
uncertainties in the sensitivities are independent of what is accounted for in
step 2, so it is justified to add them in quadrature to obtain the total error
bar on each of the corrected $A_z$ for the 23 sets. This uncertainty is also
included in the ``systematic'' uncertainty in Table \ref{corr}.

\item The $A_z$ reported is the weighted mean of the 23 data sets with
a weight $1/err^2$, where $err$ is the ``total'' uncertainty, not including
energy modulation. The error bars shown in the bottom panel of Fig.
\ref{data} are these ``total'' uncertainties.

\item As mentioned earlier, the correction for extracted energy modulation was
complicated by the fact that no direct measurement of energy modulation could
be made in the parity beamline. The energy modulation sensitivities depended on
the beamline tune, so the distribution of measured OPPIS energy modulation
values and $dA_z/dE_{oppis}$ sensitivities was examined for a given beamline
tune,  and a correction and a ``worst case'' uncertainty in the correction were
estimated.

Net corrections were calculated for energy modulation on a year by year basis.
1998 and 1999 required two corrections each, one for each beamline helicity.
1997 needed four corrections, as two Wien Filter settings were used. Finally
all the energy modulation corrections were combined and one correction was
applied to the final $A_z$.

The energy modulation correction shown in Table \ref{corr} is the net
effect of energy modulation over the three runs.  The net correction was
zero. The $\pm 0.12$ uncertainty comes from the quadrature sum of the
uncertainty in energy modulation plus the uncertainty in the energy
modulation sensitivities. This was not included in the individual error bars
for the 23 $A_z$ numbers because the energy modulation was not known well
enough. As a result, the reduced chi squared of 1.08 for the 23 sets is
larger than it really ``should'' be if the energy modulation uncertainty
was determined individually for each of the 23 sets and included in the
individual set by set error bars.

\end{enumerate}

\begin{figure}[t]
\includegraphics[width=85mm]{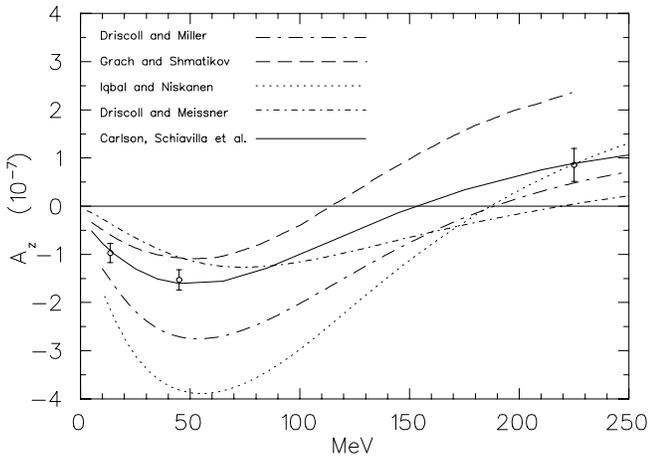}
\caption{Theoretical predictions for $A_z$ and the most precise experimental
data at 13.6 MeV (Bonn), 45 MeV (PSI) and 221 MeV (TRIUMF). The
solid curve shows the results obtained by Carlson {\em et al.} by adjusting the
weak coupling constants for the best fit to the experimental data. }
\label{theory}
\end{figure}

\section{ Results}

The overall result for $A_z$ from the 23 sets is summarized in Table
\ref{corr}. After correcting for systematic errors, the longitudinal analyzing
power is found to be $ A_z = (0.84 \pm 0.29(stat.) \pm 0.17(syst.)) \times
10^{-7}$ at 221.3 MeV incident proton energy and the target and detector
geometry of this experiment. Correcting for finite geometry and target
thickness (see Sec. \ref{subsec:energy}) and combining the errors in
quadrature, gives $A_z = 0.86 \pm 0.35$ at 225 MeV for comparison with
theoretical calculations.

Parity violation in $\vec{p}p$ scattering has already attracted considerable
theoretical interest, and many calculations of $A_z$ have been made
\cite{Carl01,Dris89,Grach93,Icqb94,Dris90}. These calculations are shown in
Fig. \ref{theory} together with the TRIUMF result (corrected to the $^1S_0 -
^3P_0$ zero crossing energy) and the most precise results at 13.6 MeV
\cite{Evers} and 45 MeV \cite{Kist87}. Theoretical predictions for $A_z$ from
several models are shown. The model of Driscoll and Miller \cite{Dris89} is
based on the Bonn potential to represent the strong N-N interaction, together
with the weak meson-nucleon coupling constants as given by Desplanques,
Donoghue, and Holstein (DDH) \cite{DDH80}. The prediction of Iqbal and Niskanen
\cite{Icqb94} has a $\Delta$ isobar contribution added to the Driscoll and
Miller model on a semi-ad-hoc basis. The theoretical prediction of Driscoll and
Meissner \cite{Dris90} is based on a self-consistent calculation, with both
weak and strong vertex functions obtained with a chiral soliton model. 
Finally, the quark model calculation of Grach and Shmatikov \cite{Grach93}
takes explicit account of quark degrees of freedom.  None of these predictions
are in good agreement with the data, although they all have similar shapes due
to the energy dependence of the strong interaction.

\begin{figure}
\includegraphics[width=85mm]{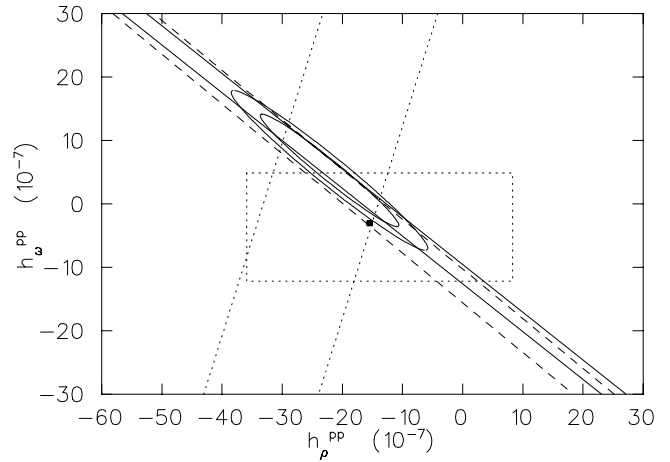}
\caption{Present constraints on the weak meson-nucleon couplings based on the
experimental data and recent calculations by Carlson et al.\cite{Carl01}. The
bands are the constraints imposed by different experiments (Bonn 13.6 MeV,
dashed; PSI 45 MeV, solid; TRIUMF 221 MeV, dotted). The filled square and
dotted rectangle are the DDH ``best guess'' and ``reasonable range''
respectively. Also shown are the 68\% and 90\% C.L.
contours. }
\label{couplings}
\end{figure}

A major source of uncertainty in these calculations is the value
of the weak meson-nucleon couplings. Starting with the work of Desplanques,
Donoghue, and Holstein (DDH) \cite{DDH80}, many theoretical calculations of
these quantities were made \cite{Dub86,Kais89,Meis90,Meis99} but
theoretical uncertainties remained large. Until the TRIUMF experiment, the
coupling constants were very poorly constrained by experiment.

Figure \ref{couplings} shows the limits on the weak meson-nucleon couplings
$h^{pp}_\rho$ and $h^{pp}_\omega$ now imposed by the low energy results
\cite{Kist87, Evers} and the present TRIUMF result. The error bands are based
on a calculation by Carlson {\em et al.} \cite{Carl01} assuming the Argonne
$v_{18}$ (AV-18) potential \cite{Wirin95}, the Bonn 2000 (CD-Bonn)
\cite{Mach01} strong interaction coupling constants, and including all partial
waves up to J=8. Although the TRIUMF measurement is not sensitive to $A_z$ from
$SP$ mixing, and the contribution from $PD$ mixing contains no
$h_{\omega}^{pp}$ contribution, there is some $h_{\omega}^{pp}$ dependence
arising from the higher partial wave mixings. The net result is that the
acceptable band defined by the TRIUMF measurement is almost orthogonal to that
defined by the low energy measurements, and greatly reduces the acceptable
ranges of both $h^{pp}_\rho$ and $h^{pp}_\omega$. Adjusting these coupling
constants for the best fit to the $pp$ data, including the TRIUMF 221 MeV
point, Carlson {\em et al.} \cite{Carl01} estimate $h^{pp}_\rho = -22.3 \times
10^{-7}$ and $h^{pp}_\omega = 5.17 \times 10^{-7}$, compared to the DDH ``best
guess'' values of $h^{pp}_\rho = -15.5 \times 10^{-7}$ and $h^{pp}_\omega =
-3.0 \times 10^{-7}$. The solid curve in Fig. \ref{theory} is calculated using
the Carlson {\em et al.} adjusted couplings.

The reduction in the experimentally allowed range for weak meson-nucleon
coupling constants also has implications for the analysis of electroweak
radiative corrections in backward angle parity violating electron
scattering. By combining back angle electron scattering data for hydrogen
\cite{Spayde00} and deuterium \cite{Hast00}, the SAMPLE collaboration was
able to extract values for both the isovector axial $e$-$N$ form factor,
$G^e_A(T=1)$, and the strange magnetic form factor, $G^s_M$ at a momentum
transfer $Q^2 = 0.1$ (GeV/c)$^2$. Their value $G^e_A(T=1)=+0.22 \pm 0.45 \pm 0.39$
differs significantly from the value $G^e_A(T=1)=-0.83 \pm 0.26$ arrived at
by Zhu {\em et al.} \cite{Zhu00} by applying one quark and many quark
(proton anapole moment) radiative corrections to the well known nucleon
axial charge as measured in neutron $\beta$-decay. $h^{pp}_\rho$ enters in
the calculation of the proton anapole moment, but its contribution is small
and one would require $h^{pp}_\rho \simeq -180 \times 10^{-7}$ to bring the
value up to the $+0.22$ of the SAMPLE measurement. Such a value of
$h^{pp}_\rho$ is now clearly ruled out by the $pp$ parity violation data.


\section{Conclusion}

The parity violating analyzing power, $A_z$, in $\vec{p}p$ elastic
scattering has been measured at 221.3 MeV incident proton energy. The result
constrains theoretical calculations of $A_z$ in an energy region not
previously covered experimentally. In the case of meson exchange
calculations, it constrains principally the value of the weak rho
meson-nucleon coupling $h^{pp}_\rho$, but, when the low energy $\vec{p}p$
data are included, strong constraints are placed on the acceptable values
of both $h^{pp}_\rho$ and $h^{pp}_\omega$. This result has implications for
the interpretation of other experiments. For example, it rules out
incorrect values of these couplings as an explanation for the disagreement
between the SAMPLE isovector axial form factor result \cite{Hast00} and the
calculation of Zhu {\em et al.} \cite{Zhu00}.

This work was supported in part by the Natural Sciences and Engineering
Research Council of Canada, the U.S. Department of Energy, and the
Russia-North America Scientific Collaboration Program; TRIUMF receives
federal funding via a contribution agreement through the National Research
Council of Canada.

\end{document}